# On the Feasibility of Deep Learning Classification from Raw Signal Data in Radiology, Ultrasonography and Electrophysiology

Szilárd ENYEDI
*Automation Department*
*Technical University of Cluj-Napoca*
Cluj-Napoca, Romania
Szilard.Enyedi@aut.utcluj.ro

*Abstract*—Medical imaging is a very useful tool in healthcare, various technologies being employed to non-invasively peek inside the human body. Deep learning with neural networks in radiology was welcome – albeit cautiously – by the radiologist community. Most of the currently deployed or researched deep learning solutions are applied on already generated images of medical scans, use the neural networks to aid in the generation of such images, or use them for identifying specific substance markers in spectrographs. This paper's author posits that if the neural networks were trained directly on the raw signals from the scanning machines, they would gain access to more nuanced information than from the already processed images, hence the training – and later, the inferences – would become more accurate. The paper presents the main current applications of deep learning in radiography, ultrasonography, and electrophysiology, and discusses whether the proposed neural network training directly on raw signals is feasible.

## I. Introduction

Each of the technologies in non-invasive medical imaging – mainly X-rays, Computed Tomography (CT), Magnetic Resonance Imaging (MRI), Positron-Emission Tomography (PET), Single-Photon Emission Computed Tomography (SPECT), Diffuse Optical Tomography (DOT), Optical Coherence Tomography (OCT) and ultrasound (US) – has its own strengths, but also its drawbacks, be it low resolution, long duration, radiation exposure, high cost, or other. Figure 1 depicts the main medical imaging technologies.

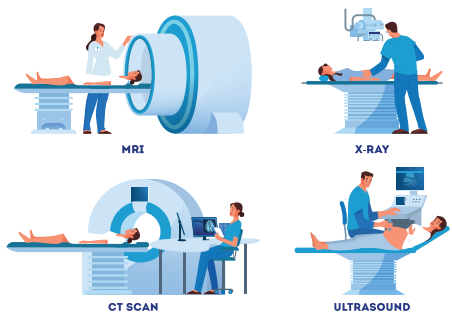

Fig. 1. Major medical imaging technologies [1]

Therefore, the medical community is interested in new solutions, including deep learning with neural networks, and its attempts to alleviate some symptoms of said technologies.

The main differences between neural networks, machine learning and deep learning:

1) **Machine learning (ML)**: Employing algorithms, historical data, and statistics, machine learning can identify patterns in the data and make predictions. It can also improve over time. However, it needs pre-processed data fed to it by humans, in order to start. Typically, the human expert establishes the features, attributes that are common among data (images) of the same category (class), and the features that are different between classes.

Machine learning does not necessarily use neural networks, although lately, most of them do.

2) **Artificial neural networks (ANN)**: Most artificial neural networks are actually implemented in software, these days. However, instead of traditional algorithms with sequences, branching and loops, their functioning mimics the nervous systems and brains of living organisms.

The main advantage of an artificial neural network over a traditional algorithm, at least in this author's view, is the high number of parallel operations and the simplicity of these operations, in neural networks, compared to the latter. This simplicity (mainly multiplications) lends itself well to parallel processing, to implementations with multiple simple processors or a few large processors with a simple but highly repetitive, high-density structure, even to implementations on low level electronics, optics or basic biological structures – from where it was all inspired.

3) **Deep learning (DL)**: DL is a subset of machine learning, built on extensive neural networks. In deep learning, there is no need for a human expert to tell the neural network what features to look for in the data. If the task is classification, the system is given (shown) a large amount of data (large number of images or other data) of a kind (labelled with that class' name), and it detects, by itself, what the common features are in those data (images). Then it is given another large set of data, all of a second kind, and it will "learn" the common features of that type of data, but it will also learn the features that are different between the first kind and the second kind (different classes). After this training phase, it can be given new, never seen data (images), and it will choose the most probable label for it, by itself. In other, non-classification tasks, DL can predict the datasets' evolution, based on old data that it was trained on.

Deep learning takes longer time and much more training data, than machine learning. The neural networks used for deep learning are also more complex than those used in machine learning, they have tens of layers, sometimes hundreds. On the other hand, it is able to find patterns – differentiating or common features – in the data, which even the best human experts might have overlooked.

Most of the currently deployed or currently researched techniques with deep learning applied to medical scans, do so by running neural networks on images of the human insides traditionally scanned with various technologies, or use the networks to generate these images, or use deep learning for detecting specific elements in spectrographs. These are understandable goals, since it is the collective experience of physicians, their colleagues, the physicians before them, and seeing with their own eyes, that drives the diagnoses.

However, during the process of image generation, some of the information gathered from the machines is discarded, quantized, or otherwise lost, in the pursuit to synthesize the visual information in an image presentable to the specialist with a trained eye. This paper's author proposes that if the neural networks were trained directly on the raw data representing the signals gathered from the scanning machine, they would gain access to the nuances in the physical data generated by the scanning process, hence the training – and later, the inferences – would become more precise. The paper presents the current applications of deep learning in radiography, with emphasis on MRI, and ponders whether such direct-signal neural network training is already or will be feasible in the near future. Figure 2 exemplifies an MRI scanner.

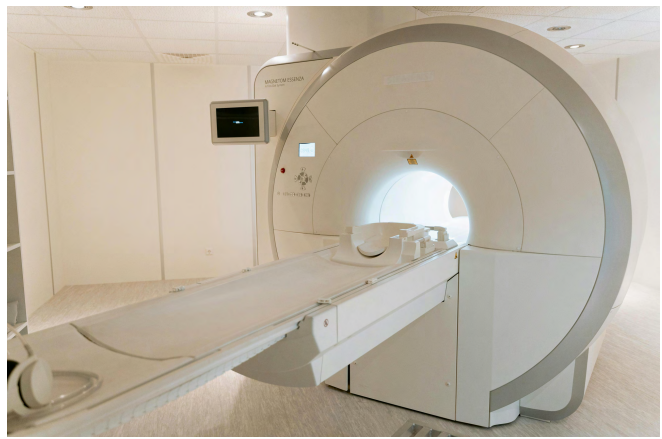

Fig. 2. Siemens Magnetom Essenza MRI machine [2] [3]

## II. Deep learning in medical imaging

### A. Feature extraction from images, and classification

This is the most prevalent use of neural networks in medical imaging [4] [5] [6] [7] [8] [9] [10] [11] [12], the main positive and negative reasons being:

1) **Automation**: For the radiologist going through the images of the scans, automating as much of the process as possible, especially pre-classifying the images (e.g. healthy/benign/malignant) once generated, but before they reach the human expert, is a great time saver and, ultimately, enables the specialist to help more people.

2) **Availability**: Neural network methods for image processing, software libraries, and neural networks pre-trained on common (non-medical) images, for transfer learning, are well established and widely used, also experts in this field are readily available.

3) **Trustworthiness**: Healthcare for humans is one of the most critical and regulated industries in the world, therefore new solutions, even additions or changes to existing methods, are regarded with high scepticism, until extensively tested and proven.

One difficulty of using deep learning to process medical images is the relative scarcity of such images for training the networks. Compared to the abundance of common images that humans snap and upload every second every day, medical images have at least an order of magnitude lower availability in numbers, for the deep learning experts and their networks to train on. Frequently, transfer learning is used – neural networks trained on other types of images, and somewhat adjusted/re-trained with specific medical image sets –, or the final layer of the network is swapped out for different, often non-neural steps – support vector machines, decision trees or other methods [13].

Another problem with sourcing images for neural network training is the inconsistency of the image data, despite the standardization brought by DICOM (Digital Imaging and Communications in Medicine). The Picture Archiving and Viewing Systems (PACS) within healthcare establishments have difficulties labelling, cataloguing, organising and presenting the relevant data to the neural networks for processing, and that is before the main ailment of big data – sheer size – comes into play. Ironically, the scarcity of images is still keeping the systems from being overwhelmed.

There are minute variations in patient positions during scans, differences in patients' anatomies, even in scans of the same patient – this is a drawback for classic image comparisons, but can be a blessing for neural network training, providing natural data augmentation.

Still, even though imaging has a proven potential to predict patient evolution, worldwide it is underutilized in guiding oncological treatment.

One group of researchers is trying to raise awareness related to "off-label" use of data [14], i.e. public datasets created and optimized for one purpose, but used and sometimes even reconstructed by other researchers as training data for another purpose, in their neural networks. This often leads to overly optimistic results, which might seem good for research, but it pushes the applicability further back from the real world, into the theoretical domain. This is another example of information loss due to the lossy processing, storage and transmission in the diagnosis chain.

### B. Image synthesis, from collected data

This field of deep learning applications places the neural networks within the pipeline of image generation. The goal is still the creation of images for the doctor to analyse, however there are several reasons for the inclusion of deep learning:

1) **Faster image generation**: Using neural networks to facilitate processing the high quantity of data coming from the machines, especially from magnetic resonance machines, makes the MRIs available faster, hence more patients can be scanned. One drawback of MRI is the lower speed of getting the results, compared to, say CT scans, and the scanners are always overbooked, even with rotating teams and multiple work shifts.

2) **Shorter exposure:** Another problem with most technologies in this field, is motion blur. Heartbeats, breathing, circulation cannot be stopped and held, for a good picture. Therefore, the shorter the scanning time, the clearer the resulting image. On the other hand, this reduces the number of details that the system is able to capture.

On the flipside, Dynamic Contrast-Enhanced (DCE) [15] technologies with contrast agents, and motion MRI [16], are specifically tuned to capture dynamic images of a living organism.

PET is molecular imaging and can yield information also at cellular level, nevertheless it exposes the body to harmful radiation, as does CT. Thus, they also benefit both from shorter scan times – shorter exposure – and reduced power scans – lower radiation dose. These can be obtained, if the quickly scanned, sparser data can be augmented with neural networks to present full detail images.

3) **More detailed images**: Even if MRI scans are generally more detailed than CT ones – although they each have their advantages and are often used together –, more details can catch more tissue abnormalities. So-called "super-resolution", neural network enhanced images, or neural network denoised images are helpful in medical imaging [17].

However, the bias has to be always on the cautious side, since these artificially enhanced images from generative neural networks [18] might erase important details that are there in the patient and might indicate illness, or create bad details that aren't really there and waste resources or result in misdiagnosis and mistreatment [19].

*C. Deep learning with spectrographs*

Specifically Nuclear Magnetic Resonance (NMR) spectroscopy can be a data and processing intensive endeavour, even if the goal is specific for the searched substance. Using neural networks to simplify the workload gives results quicker, especially if there is a sufficiently large and diverse training data set. The ongoing research shows that it is a promising direction.

Due to the connection between cancer and changes in metabolism, NMR is useful in this research field. A team in Brazil added NMR spectroscopy on blood to their arsenal, combined it with other investigations, ran them through machine learning and managed to obtain 80% accuracy in predicting whether breast cancer patients will resist neoadjuvant chemotherapy [20].

*D. Deep learning in radiogenomics*

Radiogenomics, linking genotypes with medical imaging phenotypes, is especially useful in predicting response evolution of cancer treatment. It is less expensive than traditional genetic sequencing, and it is also less intrusive, reducing biopsies. There is research taking deep learning to radiogenomics [21] [22]. Unfortunately, radiogenomics is also very complex, therefore it is still more in the research domain, rather than in the radiologist's everyday practice. Neural networks and deep learning can take the burden off classic algorithms, computing hardware and the specialists, leading to a wider adoption of this solution in radiogenomics.

*E. Ultrasound and neural networks*

Putting deep learning in the hands of sonographers and sonologists seems to be more difficult than for radiologists [23] [24]. The big advantage of ultrasonography with hand-held transducers – real-time, customized yet inexpensive imaging – also makes it less consistent in its data, therefore in training data for neural networks.

1) **Cable thickness:** A pressing aspect of modern medical ultrasonography is cable capacity, which hardly keeps up with the data flow. The cable linking the handheld probe to the main machine must be nimble, to allow free motion for the healthcare professional in exploring the subject. However, the quantity of data gathered from a modern ultrasound head is very high, not least due to the modern ultrasound scanner's frequencies, ranging from 2 MHz to 15 MHz [25]. The high frequencies allow for more details and spatial resolution, but they are also more absorbed by various tissues, thus the wide range of frequencies utilized, to be adapted depending on the examined organ and tissue.

2) **Probe size:** The bandwidth limitation is exacerbated by catheter ultrasound transducers, where the thinner the cable, the better. Therefore, the system's designers must balance gathering as much raw data as possible from the transducer, the bandwidth for real-time transmission of this data, and the physical limitations for the ultrasound procedure. Figure 3 shows a catheter probe.

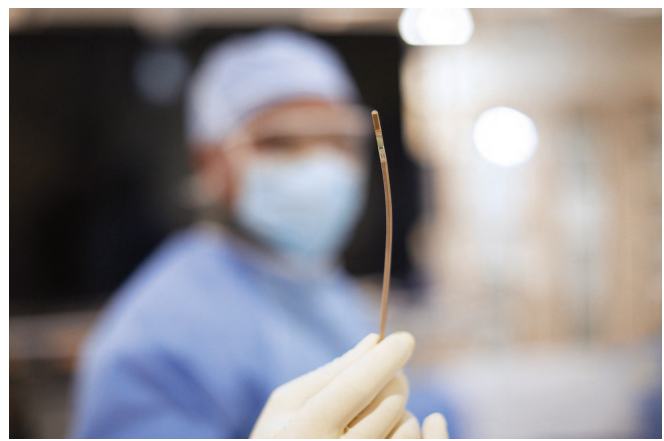

Fig. 3. Philips Verisight intra-cardiac echocardiography probe [26]

Hence, one of the most dynamic areas of research in this field is making the probe itself as small as possible [27], and the connections between the probe and the main ultrasound machine as thin as possible [28], and as fast as possible:

3) **Probe and cable reduction:** For the cable, one solution is to do the signal filtering and pre-processing in the probe head, so that raw data does not have to be transported through the cable. Compression is routinely used, but further economy is welcome. This is where not only traditional algorithms, but neural networks can help – deciding over and sending only the important data, or reconstructing data from a lesser source [29].

For very small probe sizes, there are various initiatives with Application-Specific Integrated Circuits (ASICs), which integrate sender/receiver/encoding mechanisms. The designers could add neural processing optimised functions to these circuits, but collecting the raw data would be best.

4) **Adaptive beamforming:** Born from the flexible nature of handheld ultrasound scanning, there is much variability, which can be exploited by the expert to have a better picture. On the other hand, the need for extensive professional experience can be reduced, if the beamforming [30], or at least part of it, is handled by trained neural networks. This allows the procedure to be ran by sonographers with less experience, thus widening the societal scope of this healthcare component.

5) **Enhanced imaging:** Similarly to MRI/tomography, images gathered by ultrasound scans can be enhanced by neural networks trained for that purpose, on those parts of the body. Still, compared to the aforementioned techniques, handheld ultrasound images are more heterogeneous due to their spatial variability. Consequently, they are less suitable for training the networks, or they need to be more numerous, for the training to recognize common patterns.

*F. Deep learning from electrograms*

The simpler the data, the easier it is to train neural networks with it. There is research in the field of electrocardiograms (ECG/EKG) [31] [32] [33] and electroencephalograms (EEG), extendable to electromyograms (EMG) and electrooculograms (EOG). For example, a Brazilian team has reached 99% specificity for 6 types of abnormalities in 12-lead ECG sessions [34].

Interestingly, a major problem also here is the lack of training data. Not because of the rarity of ECG or EEG sessions, but that of recorded raw data available for training. Even though the importance of the "end-to-end" (from machine to pre-diagnosis) chain is recognized, research is now looking at the opposite of the problem radiology has: sourcing current and historical paper-based/digitised ECGs/EEGs and training neural networks on those images might be more at hand than sourcing raw ECG signal data, although the latter arguably gives more solid results in deep learning [35].

III. MANAGING RAW DATA

Since its introduction in 1993, the DICOM format has become the standard for medical images, allowing for the digitalization of this part of the industry, and vastly improving the exchange of such information between healthcare participants.

And yet, CT, MRI and other specialists often want to save and/or share (in anonymized form) their data, all data, not just the final, processed DICOM images. A typical CT scan yields a few tens of megabytes of images in DICOM format, even if compressed. MRI, X-rays or other methods are comparable. Whereas, MRI k-space data can take 4-5 GB per MP2RAGEME (magnetization-prepared 2 rapid acquisition gradient echo, multi-echo extension) sequence [36], and a whole scan about 25 GB per subject. This is much larger – and more detailed – than the standard DICOM image set created by the MRI machine's hardware and software suite, and the operator. However, they are also typically proprietary and lose interoperability with common viewing software, a big disadvantage for the medical world accustomed to DICOM. One solution is to save the data in simple CSV (comma-separated values) or TSV (tab-separated values) files, however this still lacks interoperability.

To mitigate this deficiency, there are several raw data format initiatives, mostly specific to either the scanning technology or the area which is scanned. Some of them are BIDS (Brain Imaging Data Structure) [37], NIfTI (Neuroimaging Informatics Technology Initiative) [38], ISMRMRD (International Society for Magnetic Resonance in Medicine Raw Data) [39], UFF (Ultrasound File Format) [40] [41], but there are other proposals, as well [42] [43] [44]. Various dataset repositories use several formats [45] [46].

Even with the file format problem solved, the storage and transmission of such large files is still not trivial, some of the hosting sites offering advice and scripts for successfully getting the datasets.

IV. DISCUSSION

Deep learning is gaining wider adoption, even in the change-resistant medical community. Image analysis, image generation and enhancement are already using it, more and more solutions will rely on deep learning as an extra crutch for better diagnosis, not only as an afterthought in software processing of images, but as standard processing, then within the scanner itself. Figures 4 and 5 depict the current process and the proposed enhancements.

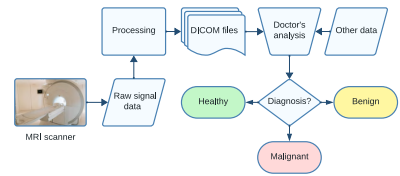

Fig. 4. Traditional MRI assisted diagnosis flow (simplified)

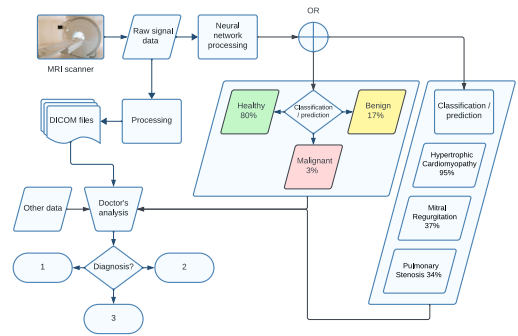

Fig. 5. Proposed diagnosis enhancements

Unfortunately, designing the neural networks for these amounts of raw signal data, hosting them, training them, is orders of magnitudes more difficult.

As summarized in the previous sections, the know-how is there. It is the poor availability of raw signal data, the storage and memory needs, and the insufficient processing power, that makes it currently unfeasible to deep train a neural network, with 5-25 GB per item in the training data set, as mentioned before. This is further aggravated by the general rule of needing 10 labelled training examples per input feature [47], of which there are many in a raw MRI machine scan. Training on big data is practiced, is being researched [48] [49] [50], and is technically possible today, but is the extra nuance in the raw signal data worth the data sourcing and training effort today?

TABLE I. presents the top five barriers against deep learning using raw signal data from the machines, in decreasing order of perceived importance.

TABLE I. BARRIERS AND SOLUTIONS

| Priority | Limitation | Possible mitigation |
|---|---|---|
| 1 | Raw signal data is scarce | • Incentivise CT/MRI/PET/SPECT/ultrasound operators, and equipment makers, to export and share their raw data.<br>• More collaboration among research/healthcare facilities and researchers<br>• Build more CT/MRI/PET/SPECT scanners, do more scans [51]:<br>  ○ Portable scanners [52]<br>  ○ Low power scanners |
| 2 | Processing power is lacking | • Training on computer grid<br>• Training on GPU (Graphical Processing Unit) farm<br>• Training on NPU farm<br>• Distributed training<br>• Training in the background<br>• Online training (while gathering new data and adapting; it takes more time, but also possibly fewer instantaneous resources)<br>• Transfer learning+online training<br>• Training on quantum computers |
| 3 | Nonstandard data formats | • Pick the most used format and make it de facto.<br>• Get CT/MRI/PET/SPECT/ultrasound/other imaging vendors and major clients to talk and establish consensus for one or a few standards. |
| 4 | Patient privacy versus sharing their scans | • Raw signal files are more anonymous than DICOM images, they contain mostly measurements of physical phenomena. |
| 5 | Lukewarm interest from specialists | • Incentivise organizations and researchers (both in medicine and neural networks) to look for funding resources, build on previous results and future brainpower, not excluding artificial intelligence. |

Luckily, more and more new processors of all kinds – from supercomputers to low power microcontrollers – contain NPUs (Neural Processing Units), beside the main processor. It is true that these are mainly designed not for training neural networks (resource-intensive), but for running inferences (execute neural network processing, easy on resources) with neural networks that have already been trained on separate, more powerful, dedicated machines. However, most of these NPUs are general-purpose enough that the differences between training the network and running the trained network are not unsurmountable, and the world's neural network training capacity is increasing.

Given enough training resources, the best fitting neural networks are born out of training from scratch. Nevertheless, to shorten the training time and resource consumption, transfer learning can be used, from the neural networks already trained and used for enhanced image synthesis, or from those used for chemical marker identification. However, most of those networks were trained for simple image generation, not classification based on big data directly from the machine.

V. CONCLUSION

This paper presents the main deep learning applications in medical imaging today, and the most promising research tendencies. The analysed literature (not all presented here) does not mention deep learning solutions directly from radiology scanner data, which could work as machine generated, recommendation value medical diagnoses, or at least disorder classification with attached probabilities. Most deep learning tasks currently work on medical images or their generation, on filtering/prioritizing data to be transmitted, or identifying specific substances/mixtures of substances from spectroscopy.

Feeding raw signal data, instead of already constructed images, into neural networks, together with their labelled disorder or medical abnormality, possibly multiple disorders or irregularities labelled by experienced specialists based on historical records, could help with better automatic machine pre-diagnoses, the doctor still having the final word.

This training is not yet feasible today, but with more raw data, the proliferation of processors optimized for neural networks, non-silicon neural networks, quantum computers with sufficient qubits being stable enough to be useful, and quantum structures and algorithms matched to neural network training [53] [54] [55] [56], it will be implemented, perhaps not that far in the future.